\documentclass[prl,aps,tightenlines,superscriptaddress,]{revtex4}
\usepackage{epsfig,amssymb,amsmath,bm}
\begin{document}

\title{Superpositions of SU(3) coherent states via a nonlinear evolution}
\author{K.\ Nemoto$^1$\cite{kaeemail} and B.\ C.\ Sanders$^2$}
\address{$^1$Centre for Quantum Computer Technology,
The University of Queensland, Queensland 4072, Australia\\
$^2$Department of Physics, Macquarie University, Sydney,
New South Wales 2109, Australia}
\date{\today}

\begin{abstract}
We show that a nonlinear Hamiltonian evolution can transform
an SU(3) coherent state into a superposition of
distinct SU(3) coherent states,
with a superposition of two SU(2)
coherent states presented as a special case.
A phase space representation is depicted by projecting
the multi--dimensional $Q$--symbol for the state to a spherical subdomain
of the coset space.
We discuss realizations of this nonlinear evolution in the
contexts of nonlinear optics and Bose--Einstein condensates.
\end{abstract}
\maketitle

\section{Introduction}
The superposition principle in quantum physics implies that
superpositions of probability amplitudes for classical--like
states are possible, yet superpositions of quasi-classical
quantum field states
are not generally observed in practice.
Such superposition states are especially interesting as they
dramatically illustrate the quantum superposition principle
by creating coherent superpositions of distinct states, with
these distinct states behaving as classical physical states.
Extensive theoretical and experimental studies have been directed
towards understanding and obtaining superpositions of distinct
quasi-classical states, often referred to as
Schr\"{o}dinger cat states\cite{Yur86a,Buz95}.
In general the interest has been in generating superpositions of
Heisenberg--Weyl coherent states via a nonlinear Hamiltonian evolution
\cite{Yur86a,Mil86}, where the Heisenberg--Weyl coherent states are the displaced
harmonic--oscillator vacuum state.

Heisenberg--Weyl coherent states are relevant for studying the
quantum--classical transition for the harmonic oscillator with
one degree of freedom.
The generalization to harmonic oscillators with two or more
degrees of freedom leads to superpositions of product
coherent states, or entangled coherent states,
with a wealth of phenomena to be studied for such systems\cite{San92}.
However, another generalization of superpositions of
coherent states is possible by employing the generalized
coherent states for general group actions\cite{Per86}.
Examples include superpositions of SU(2) coherent states\cite{San89}
and superpositions of SU(1,1) coherent states\cite{Ger87},
as well as superpositions of multiparticle SU(2) and
SU(1,1) coherent states\cite{Wan00}.
Whereas the Heisenberg--Weyl group is the symmetry group for
harmonic oscillators, SU(2) is the symmetry group for spin
precession, for two--channel interferometry and for the dynamics of ideal
two--level atoms\cite{Per86,Rad71,Are72,Yur86b},
and SU(1,1) is the symmetry group for the production of
squeezed light and quantum interferometry with parametric
up-- and down--conversion\cite{Ger87,Yur86b}.
The generalized coherent states for such systems represent
the quasi-classical states.
For example the SU(2) coherent states are also known
as atomic coherent states and are analogous to classical electric
dipoles\cite{Are72}.
Superpositions of these generalized coherent states
in quantum systems are therefore
of interest in studying counter-intuitive manifestations of
the superposition principle.

The studies of superpositions of Heisenberg--Weyl, SU(2) and SU(1,1)
coherent states are simplified by the fact that these are
one--parameter groups, and the corresponding coset space is thus a
locally two--dimensional manifold.  For the Heisenberg--Weyl group, the
manifold is the plane, for SU(2) the Poincar\'{e} sphere and for
SU(1,1) the Lobachevsky plane.  Here we generalize the studies of
superpositions of coherent states to a two--parameter Lie group, namely
SU(3).  The manifold for phase space dynamics is the four--dimensional
space isomorphic to the coset space SU(3)/U(2).  Working in a manifold
which is greater than two dimensions presents problems in terms of
visualizing the dynamics, which we resolve by projecting the dynamics
to subdomains of spherical manifolds.

In order to generate superpositions of SU(3) coherent states,
we consider the three--dimensional nonlinear oscillator.
Such a model can be realized by employing
the three--boson realization of the SU(3) generators
and constructing the dynamics optically through employing passive
optical devices and Kerr--type nonlinearities.
Another physical system involves three interacting, independent
Bose--Einstein condensates (BECs) \cite{Nem00a}.
In order to bring out the essential properties of superpositions
of SU(3) coherent states, we treat the dynamics as a closed system.

Superpositions of SU(3) coherent states may be interesting from the
perspective of weak force detection and accurate measurements.  For
example, weak force detection for two coupled, separated BECs employs a
superposition of two extremal SU(2) coherent states\cite{Cor99}.
Superpositions of two extremal SU(2) coherent states are valuable for
related applications, such as precision measurement in particle
interferometry\cite{Bol96}.  Symmetries of SU(3) coherent states may
also prove to be valuable in the context of weak force detection for
three coupled BECs or for three--channel particle interferometry, as
examples.

We show that an initial SU(3) coherent state in the nonlinear oscillator
evolves to a superposition of SU(3) coherent states.
An analytical solution is provided which shows explicitly that
a superposition of sixteen SU(3) coherent states occurs.
The analytical solutions for the SU(3) case are more complex and interesting than those for SU(2),
and we elaborate on the superposition of two SU(2)
coherent states as a special case.
Graphical solutions are employed by using the $Q$--symbol (sometimes
referred to as the $Q$--function) for
the state and projecting to spherical subdomains.
The graphical method provides an intuitive picture of the dynamics.

\section{Three--dimensional nonlinear oscillator}
\label{sec:three}

The Hamiltonian for the three-dimensional nonlinear oscillator
includes three pairs of annihilation and creation operators,
designated by
$\hat c_1, \hat c_1^\dagger, \hat c_2, \hat c_2^\dagger,
    \hat c_3, \hat c_3^\dagger$,
with each pair $\{ c_i, c_i^\dagger \}$ corresponding to one degree of freedom, or ``mode''.
The usual boson commutation relation
$[\hat{c}_k,\hat{c}_j^{\dagger}]=\delta_{kj}$ applies.
We also define number operators as $\hat{n}_k=c_k^{\dagger}c_k$,
and a total number operator $\hat{N}=\sum^3_{k=1}\hat{n}_k$.
We consider an isolated system;
hence, total energy and total particle number are conserved.
In particular we are interested in the simplest system with
energy and number conservation which will allow an SU(3)
coherent state to evolve into a superposition of SU(3) coherent states.
Such a Hamiltonian is given by
\begin{equation}
\label{H:nonlinear}
H = \omega\sum^3_{k=1} \hat{n}_k
    + \chi_1(t) \sum^3_{k=1}\hat{n}_k^{\dagger}(\hat{n}_k-1)
    + \chi_2(t) \sum^3_{\stackrel{j,k=1}{j\neq k}}
    \hat n_k \hat n_j + \sum_{\stackrel{j,k=1}{j\neq k}}^3
	\Omega_{jk}(t) c_j^\dagger c_k~,
\end{equation}
with~$\Omega_{jk}(t)=\Omega^*_{kj}(t)$.
In this model, the coupling constant~$\chi_1$ corresponds to
the strength of the self-modulation term,
and $\chi_2$ is the strength of the cross--modulation term.
The time--dependent quantities~$\chi_i$ are assumed to be
controllable as well as the
time--dependent coupling strengths~$\Omega_{jk}$ that are responsible
for linear coupling between modes.  For the case that each mode
corresponds to bosons in a particular spatial region, the
terms~$\chi_i$ quantify the nonlinear interactions, and $\Omega_{jk}$
quantify the strength of (linear) quantum tunneling of bosons between
modes.  For $\chi_i=0$, with $i=1,2$, the
Hamiltonian~(\ref{H:nonlinear}) is a linear combination of su(3)
generators, and thus the evolution is linear.

The Hamiltonian~(\ref{H:nonlinear}) corresponds to two potential
physical realizations, the first being a nonlinear optical four--wave
mixing medium, that is a medium with a $\chi^{(3)}$ nonlinearity.  In
four--wave mixing, the three boson operators~$c_k$ correspond to
single--mode fields interacting in the medium, and the
coefficients~$\chi_i$, $i \in \{ 1,2 \}$ are obtained by judiciously
constructing the $\chi^{(3)}$ tensor coefficients.

A second realization arises in the context of interacting
BECs with a small number of
particles\cite{Nem00a,Ger97}, which is of interest for some condensates
\cite{Sac98}.  The three boson operators correspond to localized
BEC modes with some overlap between the modes (in contrast to the case
considered in Ref.~\cite{Nem00a}, which assumes a negligible overlap).
The localized modes are centered around the three minima of a
three-dimensional trapping potential for the BEC.
The nonlinear interactions arise due to atomic collisions and
conserve boson number.  The coefficients for the nonlinear interactions
are~$\chi_1$, for self interactions, and~$\chi_2$, for intermodal
collisions.  These parameters depend on the $s$--wave scattering length
for the condensate and its density as well as the degree of overlap
between modes for~$\chi_2$.  
These coefficients of nonlinear dynamical terms are time--dependent:
the scattering length can be varied by applying an external magnetic field
to exploit a Feshbach resonance in the collisions between the
atoms\cite{Cor00}. 

The time--varying linear quantum tunneling terms for the BEC, with
coefficients~$\Omega_{jk}$, may be varied by modifying the optical
potential barrier between minima of the trapping potential.  The terms
are left on for a short time to produce the desired SU(3) coherent
state and then shut off. Of course they are not completely eliminated
as some overlap is needed for the cross--mode collisions (with
coefficient~$\chi_2$) to take place.  It is important that the
collision terms be strong enough to be nonnegligible even when linear
quantum tunneling may be ignored.  Tuning the scattering length is
advantageous to ensure that the intermodal collision term is sufficiently
strong.  Although $\chi_2$ might be weak, our subsequent analysis is
valid for any choice of $\chi_2$.

A typical initial condition could be $N$ bosons in one region or mode.
Allowing the system to evolve under the Hamiltonian~(\ref{H:nonlinear})
with~$\chi_i=0$, causes the system to evolve linearly, via coherent
quantum tunneling, to a Perelomov SU(3) coherent state\cite{Per86}.
As a special case, relevant to weak force detection\cite{Cor99} and
particle interferometry\cite{Bol96}, extremal coherent states are
states which are mapped to zero under a ladder operator.  For example,
the SU(2) coherent state $\vert j \, j \rangle$ is annihilated by the
specified raising operator $J_+$, and the state $\vert j \, -j \rangle$
is annihilated by the specified lowering operator $J_-$.  These two
states are extremal, and a superposition of these two extremal states
would be $( \vert j \, j \rangle + \vert j \, -j \rangle )/\sqrt{2}$.
Extremal SU(3) coherent states are states which can be annihilated by
the su(3) ladder operators.


From conservation of the total number $N$ of bosons,
the Hamiltonian may be rewritten in terms of the SU(3) generators and
the total number $N$ is the irreducible representation (irrep) parameter.
The eight generators of SU(3) may be defined as the two Cartan operators
\begin{mathletters}
\begin{eqnarray}
\label{generator:Cartan1}
X_1 &=& \hat{c}_1^{\dagger}\hat{c}_1-\hat{c}_2^{\dagger}\hat{c}_2, \\
\label{generator:Cartan2}
X_2 &=& \frac{1}{3}(\hat{c}_1^{\dagger}\hat{c}_1+\hat{c}_2^{\dagger}\hat{c}_2 -
2\hat{c}_3^{\dagger}\hat{c}_3^{\dagger})
\end{eqnarray}
\end{mathletters}
and the six generators
\begin{mathletters}
\begin{eqnarray}
\label{generator:otherY}
Y_k &=& i(\hat{c}_k^{\dagger}\hat{c}_j-\hat{c}_j^{\dagger}\hat{c}_k)  \\
\label{generator:otherZ}
Z_k &=&  \hat{c}_k^{\dagger}\hat{c}_j+\hat{c}_j^{\dagger}\hat{c}_k ,
\end{eqnarray}
\end{mathletters}
where $k=1,2,3$ and $j=k \bmod 3 +1$.
The raising operators and lowering operators are defined as
\begin{mathletters}
\begin{eqnarray}
J^k_+ &=& \hat{c}_k^{\dagger}\hat{c}_j, \; (k<j),\\
J^k_- &=& \hat{c}_k^{\dagger}\hat{c}_j, \; (k>j),
\end{eqnarray}
\end{mathletters}
or, alternatively, for $k<j$
\begin{mathletters}
\begin{eqnarray}
J^k_+ &=& \frac{1}{2}(Z_k-iY_k)\\
J^k_- &=& \frac{1}{2}(Z_k+iY_k),
\end{eqnarray}
\end{mathletters}
in terms of SU(3) generators.

The term involving the irrep parameter $N$ can be removed by moving to
a rotating picture.  Furthermore, by assuming that the initial SU(3)
coherent state has been generated from the
evolution~(\ref{H:nonlinear}), with large linear quantum tunneling and
negligible contributions from nonlinear evolution, the optical
potential barrier between trapping potential minima is increased to
reduce the linear tunneling.  At the same time, the nonlinear
coefficients must be reasonably large compared to the coefficient for
linear tunneling, perhaps by employing an external magnetic field to
exploit a Feshbach resonance, as discussed above.  For large
nonlinearities and small linear quantum tunneling terms, the
Hamiltonian~(\ref{H:nonlinear}) can be approximated by
\begin{equation}
\label{hamiltonian}
H = \frac{\chi}{2}(X_1^2 + 3 X_2^2),
\end{equation}
with $\chi=\chi_1-\chi_2$.  The Hamiltonian is a sum of quadratic forms
of Cartan operators which commute, and this property will be useful in
later calculations.

\section{Coherent states}
\label{sec:SU(3)}
\subsection{SU(2) coherent states}
\label{subsec:SU(2)}
In order to introduce the SU(3) coherent states,
it is useful to review the SU(2) coherent states and
to use this knowledge to generalize to SU(3) coherent states.
SU(2) coherent states and their methods were first developed as
atomic coherent states \cite{Rad71,Are72} to treat atomic systems.
On the other hand, coherent states for Lie groups were defined by
Perelomov\cite{Per86} as an orbit generated by group action on a
reference state, which is usually the highest-- (or lowest--) weight
state.  For compact groups, the simplest nontrivial case is SU(2),
which can be parametrized by three real parameters
$\{\theta,\varphi_1,\varphi_2\}$.
The lowest--order faithful representation of an arbitrary
$g \in $ SU(2) is as a
$2 \times 2$ matrix
\begin{eqnarray}
\label{su2}
  g (\varphi_1,\theta,\varphi_2) &=&
  \left( \begin{array}{cc}
       e^{i\varphi_1}\cos\theta & e^{i\varphi_2}\sin\theta \\
       -e^{-i\varphi_2}\sin\theta & e^{-i\varphi_1}\cos\theta\\
     \end{array}\right) .
\end{eqnarray}

As SU(2) $\subset$ SU(3),
the three generators of SU(2) can be obtained from
(\ref{generator:Cartan1}), (\ref{generator:Cartan2}),
(\ref{generator:otherY}) and (\ref{generator:otherZ}).
For example, the subgroup SU(2)$_{12}$ is generated
by $\{ X_1,Y_1,Z_1\}$.
The Casimir invariant is $J^2=X_1^2+Y_1^2+Z_1^2$
with eigenvalue $j(j+1)$.
The $2 \times 2$ representation corresponds to $j=1/2$.
The weight basis corresponds to $\vert j m \rangle$
with $J_z \vert j m \rangle = m \vert j m \rangle$.
The highest--weight state is $\vert j j \rangle$
for which $J_+ \vert j j \rangle = 0$.

The SU(2) coherent state, for fixed $j$, is given by
\begin{equation}
\vert \theta , \varphi \rangle
    = \exp \left[ -\frac{\theta}{2}
        \left( J_+^1 e^{-i\varphi} - J_-^1 e^{i\varphi} \right)
            \right] \vert j \, j \rangle .
\end{equation}
The coherent state can be represented geometrically on
the Poincar\'e sphere, which is isomorphic to the
coset space SU(2)/U(1).

\subsection{SU(3) coherent states}

The SU(3) coherent states may be formulated as a generalization of the
SU(2) coherent states\cite{Gnu98,Nem00b}, and are obtained by the action
of SU(3) on the highest--weight state.  In order to obtain the SU(3)
coherent states, it is convenient to employ the decomposition of SU(3)
in Ref.~\cite{Row99}, whereby the SU(3) operator is decomposed into a
combination of SU(2) operators.  The decomposition allows us to
parameterize an arbitrary element $g$ in the $3 \times 3$ matrix
representation as
\begin{equation}
\label{su_3 matrix}
g = \left( \begin{array}{ccc}
       1 & 0 & 0\\
       0 & & \\
       0 & & V
     \end{array}\right)
    \left( \begin{array}{ccc}
        e^{i\varphi} \cos{\theta} & -\sin{\theta} & 0 \\
        \sin{\theta} & e^{-i\varphi} \cos{\theta} & 0\\
        0 & 0 & 1\\
     \end{array} \right)
    \left( \begin{array}{ccc}
       1 & 0  & 0\\
       0 & \; & \;\\
       0 & & W \;
     \end{array}\right) ,
\end{equation}
where $V(\varphi_1,\xi,\varphi_2)$ and $W(\varphi_3,\zeta,\varphi_4)$
are $2\times 2$ matrices representing elements
of SU(2)$_{23}$.
The middle matrix on the right--hand side of Eq.\ (\ref{su_3 matrix})
is an element of SU(2)$_{12}$.

The SU(3) generators of Eqs.\ (\ref{generator:Cartan1},\ref{generator:Cartan2})
and (\ref{generator:otherY},\ref{generator:otherZ}) are presented as a
three--boson realization.
It is convenient to employ the basis states,
which are eigenstates of the SU(3) Casimir operators
and of the two elements of the Cartan subalgebra
(\ref{generator:Cartan1},\ref{generator:Cartan2}).
The basis state can be labeled by the three
numbers $n_1$, $n_2$ and $n_3$, which satisfy
\begin{equation}
\hat{n}_k|n_1,n_2,n_3\rangle=n_k|n_1,n_2,n_3\rangle,
    \;\mbox{for }k \in \{ 1,2,3 \}~.
\end{equation}
As any action of the raising operators on the state $\vert N,0,0\rangle$
annihilates it, the state $\vert N,0,0\rangle$ is a highest--weight state.
Coherent states are obtained by SU(3) action on this highest--weight state.

Taking the highest--weight state $\vert N,0,0\rangle$ as the reference state,
the action of Eq. (\ref{su_3 matrix}) on the highest--weight state gives 
an explicit form of coherent states.
The factorization (\ref{su_3 matrix}) ensures that the matrix on
the furthest right leaves the highest--weight state invariant,
and only the two other matrices are important in determining the coherent state.
The highest--weight state is thus invariant under SU(2)$_{23}$;
hence the coherent states are parametrized on the coset
space SU(3)/SU(2).
In fact the SU(3) coherent states can be parametrized
on the coset space SU(3)/U(2) by eliminating
an arbitrary phase, but the following calculations
are easier if we retain the phase.
In calculations of the $Q$--symbol for the state,
this phase is not important.

The coherent states can be generated as
\begin{equation}
\vert \xi, \theta, \varphi,\varphi_1,\varphi_2\rangle
= g(\varphi_1,\xi,\varphi_2,\varphi,\theta,
\varphi_3,\zeta,\varphi_4)\vert N,0,0\rangle ,
\end{equation}
which can be expressed as
\begin{eqnarray} \label{cs}
\vert \xi, \theta, \varphi, \varphi_1, \varphi_2\rangle
    =\sum^{N}_{j_1=0} & & e^{i\varphi(N-j_1)}\sin^{j_1}(\theta)\; \cos^{(N-j_1)}({\theta})
    \left( \begin{array}{c}
    N\\     j_1\\
    \end{array}\right)^{1/2}
        \nonumber   \\
&&\times  \sum^{j_1}_{j_2=0}e^{ij_2\varphi_2}e^{i\varphi_1(j_1-j_2)}\sin^{j_2}(\xi)  \;
    \cos^{(j_1-j_2)}({\xi})
    \left( \begin{array}{c}
    j_1\\   j_2
    \end{array}\right)^{1/2}
    \vert N-j_1,j_1-j_2,j_2 \rangle.
\end{eqnarray}
This parameterization of the SU(3) coherent state includes an arbitrary phase.

\section{Quantum dynamics of the system}
\subsection{Analytical solutions}

Let the initial state~$|\psi (0)\rangle$ be an arbitrary SU(3) coherent state.
The state $|\psi(t)\rangle$ for
arbitrary time $t$ may thus be given as
\begin{eqnarray}
\label{solution}
|\psi (t)\rangle =&& \sum^N_{j_1=0}e^{i\varphi (N-j_1)} \sin^{j_1}(\theta) \;
        \cos^{(N-j_1)}(\theta)
        \left( \begin{array}{c}
            N\\
            j_1\\
        \end{array}\right)^{1/2}
    \sum^{j_1}_{j_2=0}e^{ij_2\varphi_2} e^{i\varphi_1 (j_1-j_2)} \sin^{j_2}(\xi)  \;
    \cos^{(j_1-j_2)}(\xi)
    \left( \begin{array}{c}
        j_1\\
        j_2\\
    \end{array}\right)^{1/2}      \nonumber\\
&\;&\;\times    \exp \left[ -2i\chi t \left( N^2/3+j_1^2+j_2^2-j_1(N+j_2)\right)\right]
        \vert N-j_1,j_1-j_2,j_2 \rangle~,
\end{eqnarray}
under the Hamiltonian evolution~(\ref{hamiltonian}).
The time--dependent element of Eq.\ (\ref{solution})
exhibits the periodic evolution of the state
and suggests that the state evolves into superposition states periodically,
by analogy with the SU(2) coherent state superposition case\cite{San89},
but with additional complexity due to
the greater number of parameters and the higher dimension of the
group manifold.

From Eq.~(\ref{solution}), we observe that the recurrence time,
for which the state evolves cyclically back into the original state,
is given by $\tau \equiv \pi/\chi$ for all values of the irrep parameter $N$.
At half the recurrence time, $\tau/2=\pi/2\chi$,
the state evolves into the superposition
\begin{equation}
|\psi(\tau/2)\rangle = \frac{1}{2}e^{-i\pi N^2/3}
\left[ -\vert \xi, \theta, \varphi,\varphi_1,\varphi_2 \rangle
    +\vert \xi, \theta, \varphi, \varphi_1+\pi, \varphi_2\rangle
    +\vert \xi, \theta, \varphi,\varphi_1, \varphi_2\rangle
    +\vert \xi, \theta, \varphi,\varphi_1+\pi, \varphi_2+\pi\rangle\right],
\end{equation}
for even $N$, or
\begin{equation}
\vert \psi(\tau/2)\rangle = \frac{1}{2}e^{-i\pi N^2/3}
\left[ \vert \xi, \theta, \varphi,\varphi_1, \varphi_2 \rangle
    +\vert \xi, \theta,\varphi+\pi, \varphi_1, \varphi_2 \rangle
    +\vert \xi, \theta, \varphi,\varphi_1+\pi, \varphi_2 \rangle
    +\vert \xi, \theta, \varphi,\varphi_1,\varphi_2+\pi \rangle\right],
\end{equation}
for odd $N$.
At one--quarter of the recurrence time $\tau/4$,
the state evolves into a superposition state with the exact form depending
on the total number $N$.
There are four types of superposition states at $\tau/4$,
classified by the quantity $N \bmod 4$.
The four types are given as the following,

\begin{enumerate}
\item If $N=4n$, for~$n$ an integer,
\begin{eqnarray}\label{4n}
|&\psi&(\tau/4)\rangle = \frac{1}{4}e^{-i\pi N^2/6} \nonumber\\
&\times&    \bigg[ \vert \xi, \theta, \varphi,\varphi_1,\varphi_2 \rangle
            +\vert \xi, \theta, \varphi,\varphi_1+\pi,\varphi_2 \rangle
        +\vert \xi, \theta, \varphi,\varphi_1, \varphi_2+\pi \rangle
        +\vert \xi, \theta, \varphi, \varphi_1+\pi,\varphi_2+\pi \rangle\nonumber\\
&+&     i \bigg( -\left\vert \xi, \theta, \varphi, \varphi_1, \varphi_2-\frac{\pi}{2}
        \right\rangle
    -\vert \xi, \theta, \varphi,,\varphi_1,\varphi_2+\frac{\pi}{2} \rangle
    + \left\vert \xi, \theta, \varphi,\varphi_1
        +\frac{\pi}{2},\varphi_2-\frac{\pi}{2} \right\rangle
    -\left\vert \xi, \theta, \varphi,\varphi_1
        +\frac{\pi}{2},\varphi_2+\frac{\pi}{2} \right\rangle
\nonumber\\
&+&     \vert \xi, \theta, \varphi,\varphi_1+\frac{\pi}{2},\varphi_2+\pi \rangle
    -\vert \xi, \theta, \varphi,\varphi_1+\frac{\pi}{2},\varphi_2 \rangle
    + \vert \xi, \theta, \varphi,\varphi_1+\pi,\varphi_2-\frac{\pi}{2}\rangle
    +\vert \xi, \theta, \varphi,\varphi_1+\pi,\varphi_2+\frac{\pi}{2}\rangle
\nonumber\\
&-& \vert \xi, \theta, \varphi,\varphi_1-\frac{\pi}{2},\varphi_2-\frac{\pi}{2}\rangle
    +\vert \xi, \theta, \varphi,\varphi_1-\frac{\pi}{2},\varphi_2+\frac{\pi}{2}\rangle
    - \vert \xi, \theta, \varphi,\varphi_1-\frac{\pi}{2}, \varphi_2 \rangle
    +\vert \xi, \theta, \varphi,\varphi_1-\frac{\pi}{2},\varphi_2+\pi\rangle \bigg)\bigg].
\end{eqnarray}
\item If $N=4n+1$,
\begin{eqnarray} \label{4n+1}
|&\psi&(\pi/4\chi)\rangle = \frac{1}{4}e^{-i\pi N^2/6}\nonumber\\
&\times&\bigg[ \vert \xi, \theta, \varphi,\varphi_1+\frac{\pi}{2},\varphi_2+\frac{\pi}{2}\rangle
    +\vert \xi, \theta, \varphi,\varphi_1-\frac{\pi}{2},\varphi_2+\frac{\pi}{2}\rangle
    +\vert \xi, \theta, \varphi,\varphi_1+\frac{\pi}{2},\varphi_2-\frac{\pi}{2}\rangle
    +\vert \xi, \theta, \varphi,\varphi_1-\frac{\pi}{2},\varphi_2-\frac{\pi}{2}\rangle
\nonumber\\
&+&  i\bigg( -\vert \xi, \theta, \varphi,\varphi_1+\frac{\pi}{2},\varphi_2\rangle
    -\vert \xi, \theta, \varphi,\varphi_1 +\frac{\pi}{2},\varphi_2+\pi \rangle
    +\vert \xi, \theta, \varphi,\varphi_1+\pi,\varphi_2 \rangle
    -\vert \xi, \theta, \varphi,\varphi_1+\pi,\varphi_2+\pi \rangle
\nonumber\\
&+& \vert \xi, \theta, \varphi,\varphi_1+\pi,\varphi_2-\frac{\pi}{2}\rangle
    -\vert \xi, \theta, \varphi,\varphi_1+\pi,\varphi_2+\frac{\pi}{2}\rangle
    +\vert \xi, \theta, \varphi,\varphi_1-\frac{\pi}{2},\varphi_2\rangle
    +\vert \xi, \theta, \varphi,\varphi_1-\frac{\pi}{2},\varphi_2+\pi\rangle
\nonumber\\
&-& \vert \xi, \theta, \varphi,\varphi_1,\varphi_2 \rangle
    +\vert \xi, \theta, \varphi,\varphi_1\varphi_2+\pi\rangle
    -\vert \xi, \theta, \varphi,\varphi_1, \varphi_2+\frac{\pi}{2}\rangle
    +\vert \xi, \theta, \varphi,\varphi_1, \varphi_2-\frac{\pi}{2}\rangle \bigg)\bigg].
\end{eqnarray}
\item If $N=4n+2$,
\begin{eqnarray} \label{4n+2}
|&\psi&(\tau/4)\rangle = \frac{1}{4}e^{-i\pi N^2/6} \nonumber\\
&\times& \bigg[ \vert \xi, \theta, \varphi,\varphi_1,\varphi_2 \rangle
        +\vert \xi, \theta, \varphi,\varphi_1+\pi,\varphi_2 \rangle
    +\vert \xi, \theta, \varphi, \varphi_1, \varphi_2+\pi\rangle
    +\vert \xi, \theta, \varphi,\varphi_1+\pi,\varphi_2+\pi \rangle
\nonumber\\
&+&     i \bigg( -\vert \xi, \theta, \varphi,\varphi_1+\pi,\varphi_2+\frac{\pi}{2}\rangle
    -\vert \xi, \theta, \varphi,\varphi_1+\pi,\varphi_2-\frac{\pi}{2}\rangle
    + \vert \xi, \theta, \varphi,\varphi_1-\frac{\pi}{2},\varphi_2+\frac{\pi}{2}\rangle
    -\vert \xi, \theta, \varphi,\varphi_1-\frac{\pi}{2},\varphi_2-\frac{\pi}{2}\rangle
\nonumber\\
&+& \vert \xi, \theta, \varphi,\varphi_1-\frac{\pi}{2},\varphi_2\rangle
    -\vert \xi, \theta, \varphi,\varphi_1-\frac{\pi}{2},\varphi_2+\pi\rangle
    +\vert \xi, \theta, \varphi,\varphi_1, \varphi_2+\frac{\pi}{2}\rangle
    +\vert \xi, \theta, \varphi,\varphi_1, \varphi_2-\frac{\pi}{2}\rangle
\nonumber\\
&-& \vert \xi, \theta, \varphi,\varphi_1+\frac{\pi}{2},\varphi_2+\frac{\pi}{2}\rangle
    +\vert \xi, \theta, \varphi,\varphi_1+\frac{\pi}{2},\varphi_2-\frac{\pi}{2}\rangle
    -\vert \xi, \theta, \varphi,\varphi_1+\frac{\pi}{2},\varphi_2+\pi\rangle
    +\vert \xi, \theta, \varphi,\varphi_1+\frac{\pi}{2},\varphi_2\rangle \bigg)\bigg].
\end{eqnarray}
\item If $N=4n+3$,
\begin{eqnarray} \label{4n+3}
|&\psi&(\tau/4)\rangle = \frac{1}{4}e^{-i\pi N^2/6}\nonumber\\
&\times& \bigg[ \vert \xi, \theta, \varphi,\varphi_1-\frac{\pi}{2},\varphi_2-\frac{\pi}{2}\rangle
    +\vert \xi, \theta, \varphi,\varphi_1+\frac{\pi}{2},\varphi_2-\frac{\pi}{2}\rangle
    +\vert \xi, \theta, \varphi,\varphi_1-\frac{\pi}{2},\varphi_2+\frac{\pi}{2}\rangle
    +\vert \xi, \theta, \varphi,\varphi_1+\frac{\pi}{2},\varphi_2+\frac{\pi}{2}\rangle
\nonumber\\
&+&     i \bigg( -\vert \xi, \theta, \varphi,\varphi_1-\frac{\pi}{2},\varphi_2+\pi\rangle
    -\vert \xi, \theta, \varphi,\varphi_1-\frac{\pi}{2},\varphi_2 \rangle
    +\vert \xi, \theta, \varphi,\varphi_1, \varphi_2+\pi\rangle
    -\vert \xi, \theta, \varphi,\varphi_1,\varphi_2\rangle
\nonumber\\
&+& \vert \xi, \theta, \varphi,\varphi_1,\varphi_2+\frac{\pi}{2}\rangle
    -\vert \xi, \theta, \varphi,\varphi_1 \varphi_2-\frac{\pi}{2}\rangle
    +\vert \xi, \theta, \varphi,\varphi_1+\frac{\pi}{2},\varphi_2+\pi\rangle
    +\vert \xi, \theta, \varphi,\varphi_1+\frac{\pi}{2},\varphi_2\rangle
\nonumber\\
&-& \vert \xi, \theta, \varphi,\varphi_1+\pi,\varphi_2+\pi\rangle
    +\vert \xi, \theta, \varphi,\varphi_1+\pi,\varphi_2\rangle
    -\vert \xi, \theta, \varphi,\varphi_1+\pi,\varphi_2-\frac{\pi}{2}\rangle
    +\vert \xi, \theta, \varphi,\varphi_1+\pi,\varphi_2+\frac{\pi}{2}\rangle \bigg)\bigg].
\end{eqnarray}
\end{enumerate}

\subsection{The $Q$--symbol}

In the previous section we have obtained analytical expressions of $\vert\psi(t)\rangle$ for
two particular time values, $\tau/2$ and $\tau/4$.
More generally however we can obtain the state analytically for all values of $t$.
The density matrix $\rho(t)$ for the system may be written in terms of
the pure state $|\psi(t)\rangle$ as
\begin{equation}
\rho(t)=|\psi (t)\rangle\langle\psi (t)|.
\end{equation}
The $Q$--symbol for this state is given by\cite{Per86}
\begin{equation}
Q(t)=\langle \xi, \theta, \varphi,\varphi_1,\varphi_2|
\rho(t)\vert \xi, \theta, \varphi,\varphi_1,\varphi_2 \rangle.
\end{equation}
The arbitrary phase disappears in the calculation of $Q(t)$.
It is thus convenient to employ the two relative phases
$\phi_1 = \varphi_1-\varphi$ and $\phi_2 = \varphi_2-\varphi_1$.
The $Q$--symbol for the state may thus be expressed as
\begin{equation} \label{Q-all}
Q(\xi, \theta, \phi_1, \phi_2: \xi(0), \theta(0), \phi_1(0),
    \phi_2(0); t)
    = \left| \sum^N_{j_1=0}\sum^{j_1}_{j_2=0}
    S^N_{j_1,j_2}(\xi,\theta, \phi_1,\phi_2;t)\right|^2,
\end{equation}
where
\begin{eqnarray}
S^N_{j_1,j_2} &=&
e^{-ij_1(\phi_1-\phi_1(0))}\big(\sin\theta\sin\theta(0))^{j_1}
    (\cos\theta\cos\theta(0)\;\big)^{N-j_1}
    \left( \begin{array}{c}
    N\\
    j_1\\
    \end{array}\right)\nonumber\\
&\;& \; \times e^{-ij_1(\phi_2-\phi_2(0))}\big(\sin\xi\sin\xi(0))^{j_1}
    (\cos\xi\cos\xi(0)\;\big)^{N-j_1}
    \left( \begin{array}{c}
    j_1\\
    j_2\\
    \end{array}\right)
    e^{-2i\chi t(j_1^2+j_2^2-j(N+j_2))}  .
\end{eqnarray}

The $Q$--symbol depends on four periodic parameters,
and therefore a plot of the $Q$--symbol would need to be embedded
in five--dimensional space to be viewed.
However, in order to view the plot we need to reduce the
number of dimensions required for plotting the function.
A simple way to plot the $Q$--symbol is to fix one parameter,
for example by setting $\theta=\pi/2$.
This corresponds to a {\em slice} of the $Q$--symbol,
and the choice of where to slice
depends on the features to be emphasized in the plot.
A slice is a standard means for representing multi--dimensional
figures; animation can be used, for example, to present the figure by
`traveling' through a sequence of slices.  We will present an example
in the following subsection.

\subsection{An Example}
\label{subsec:example}
The nonlinear SU(2) oscillator generates
superposition states from an initial SU(2) coherent state and
can generate a superposition of two distinct SU(2) coherent states
at a half of its recurrence time \cite{San89}.
This superposition of two distinct SU(2) coherent states
results in two major peaks arising in the plot of the $Q$--symbol for the superposition state.
Interference fringes midway between the two peaks arise due to the
coherence properties of the superposition.
These interference fringes can be clearly seen as dimples
in a plot of the logarithm of the $Q$-symbol\cite{Gag95}.

The superposition states we have seen in Eqs.~(\ref{4n}-\ref{4n+3}) arise from
the nonlinearity of the SU(3) system.
In this subsection, we wish to observe a manifestation of such
a superposition by plotting the $Q$--symbol for the state.
As discussed in the previous subsection,
we can reduce the dimension of the parameter space to two by
obtaining a slice of the $Q$--symbol for the state.
We consider an SU(3) coherent state which is also an SU(2)$_{23}$ coherent state.
This restriction to SU(2)$_{23}$ coherent states ensures that only
a subset of the $Q$--symbol is explored,
with this subset being a slice which can be plotted on the Poincar\'{e}
sphere corresponding to the SU(2)$_{23}$/U(1) coset space.
An initial SU(2)$_{23}$ coherent state can be written as
\begin{eqnarray}
\label{initialcondition}
\vert \psi(0)\rangle &=& \vert \xi(0),\pi/2,\phi_1(0),\phi_2(0) \rangle\nonumber\\
        &=& e^{-i\phi_1(0) N}\sum^N_{j_2=0}e^{i\phi_2(0) j_2} \sin^{j_2}(\xi(0))
        \cos^{(N-j_2)}(\xi(0))
    \left( \begin{array}{c}
        N\\
        j_2\\
    \end{array}\right)^{1/2}
|0,N-j_2,j_2\rangle
\end{eqnarray}
A state in the SU(2)$_{23}$ subsystem stays in this subsystem under 
time evolution, so the factor of $\phi_1$ becomes an arbitrary phase
and disappears in the calculation of the $Q$--symbol. We now
employ the notation $\tilde{Q}$ to refer to a slice of the
$Q$--symbol. In this notation, the reduced $Q$--symbol, with the
initial condition~(\ref{initialcondition}), is given by
\begin{eqnarray}
\tilde{Q}(\theta, \xi,\phi_2;t)&=& \langle \xi, \theta,\phi_1,\phi_2|
    \rho(t)\vert \xi, \theta,\phi_1,\phi_2 \rangle
\nonumber \\
&=& \sin^{2N}(\theta) \sum_{p=0}^N\sum_{q=0}^Ne^{-i(p-q)(\phi_2-\phi_2(0))}
    \big(\sin(\xi)\sin(\xi(0))\big)^{p+q}\big(\cos(\xi)\cos(\xi(0))\big)^{(2N-(p+q))}  
    \nonumber\\
&&\times\left( \begin{array}{c}
        N\\
        p\\
        \end{array}\right)
        \left( \begin{array}{c}
        N\\
        q\\
    \end{array}\right)
    \exp \left[ -2i\chi t (p-q)(p-q-N) \right]~.
\end{eqnarray}
The $\theta$--component of the $Q$--symbol is time--independent
with a fixed weighting of~$\sin^{2N}(\theta)$. The optimal slice for
observing the $Q$--symbol in the reduced $\xi$--$\phi_2$ space is
thus obtained by setting $\theta=\pi/2$.

\begin{figure}
\center{ \epsfig{figure=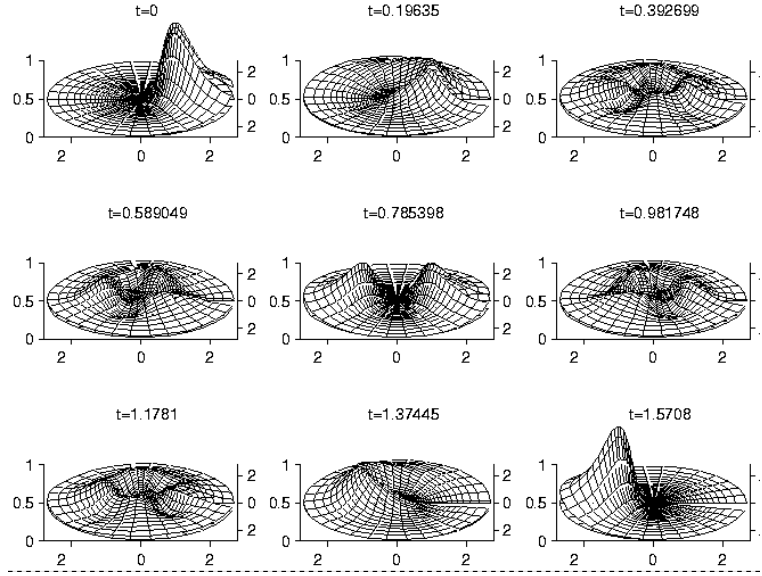,width=100mm}}
\caption{$\tilde{Q}(\xi, \phi_2)$ is plotted at a certain time $t$
with the use of the stereographical mapping.  The origin corresponds the north pole of the
spherical subdomain $(\xi, \phi_2)$.  Time flows from the left top figure at $t=0$
to the right bottom figure at $t=\tau/2$.
The recurrence time is twice as longer as the odd number case below.}
\label{f_even}
\end{figure}

\begin{figure}
\center{ \epsfig{figure=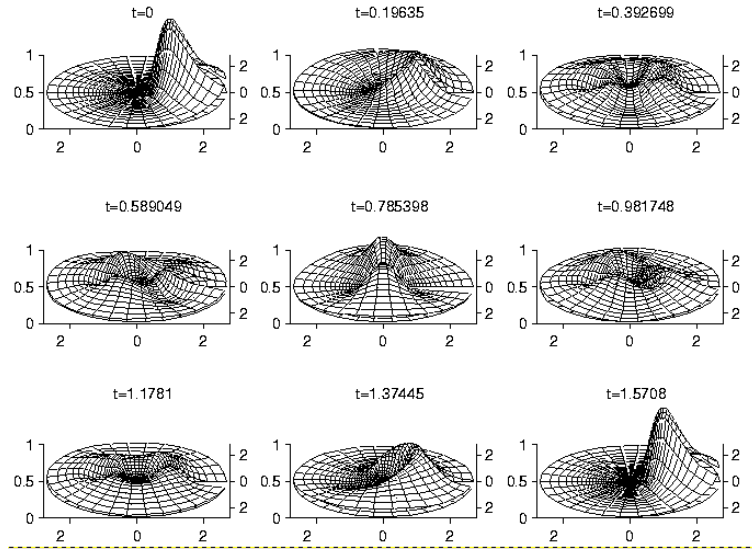,width=100mm}}
\caption{$\tilde{Q}(\xi, \phi_2)$ is plotted at a certain time $t$
with the use of the stereographical mapping.  The origin corresponds the north pole of the
spherical subdomain $(\xi, \phi_2)$.  Time flows from the left top figure at $t=0$
to the right bottom figure at $t=\tau/2$.
The recurrence time is a half of the even number case above.}
\label{f_odd}
\end{figure}

With the initial condition $\xi(0) = \pi/4$, $\phi_2(0)=0$,
$\tilde{Q}$ is simplified to
\begin{eqnarray} \label{partial-q-ex}
\tilde{Q}(\xi,\phi_2:t)
= 2^{-N}\sum_{p=0}^N\sum_{q=0}^N && e^{-i(p-q)\phi_2}
\left( \begin{array}{c}
    N\\
    p\\
    \end{array}\right)
    \left( \begin{array}{c}
    N\\ q\\
    \end{array}\right)
    \sin^{p+q}(\xi)\cos^{(2N-(p+q))}(\xi)       \nonumber   \\  &&
    \times \exp \left[ -2i\chi t (p-q)(p+q-N) \right]~,
\end{eqnarray}
and is numerically represented in Fig.~\ref{f_even} and Fig.~\ref{f_odd}.
The time--dependent term in Eq.\ (\ref{partial-q-ex})
implies that the time--evolution of the state is
distinguishable in terms of the total number $N$.  The product $(p-q)(p+q-N)$ is even for any
$p$ and $q$ if $N$ is odd,
whereas it is not true for any even total number $N$.
The recurrence time of the time evolution with an odd total number $N$ is a half of
the recurrence time with an even total number.
We solve this evolution numerically and plot $\tilde{Q}$
as it evolves for two values of the total number, $N=9,10$.
Fig.~\ref{f_even} shows
the time--evolution of $\tilde{Q}$ for the state for even number $N=10$;
in contrast, Fig.~\ref{f_odd}
depicts the time--evolution of the $\tilde{Q}$
for the state for odd number $N=9$.

\section{Conclusions}

We have investigated the formation of superpositions of SU(3) coherent
states via a nonlinear Hamiltonian evolution.  This evolution could be
realized in terms of four--wave mixing in nonlinear optics or in terms
of a nonlinear interaction between three independent modes of a
Bose--Einstein condensate with nonnegligible nonlinear interactions
between the separate, but overlapping, modes.  The linear quantum
tunneling term is allowed to vary independently of the nonlinear
(collision) terms, and linear quantum tunneling is used to prepare the
SU(3) coherent state for~$N$ bosons in the three modes.
Then this SU(3) coherent state undergoes a
nonlinear evolution into a superposition of SU(3) coherent states.

We have obtained explicit expressions for the superposition of coherent
states at one--half and one--quarter the recurrence time for this
periodic evolution. Whereas superpositions of pairs of coherent states
are obtained for half the recurrence time for Heisenberg--Weyl, SU(2)
and SU(1,1) systems, the higher dimension of SU(3) dynamics leads to
much more complex and interesting superpositions even at half the
recurrence time.  The nature of the superposition depends on the
quantity $N \bmod 4$, with $N$ being the total quantum number. The
state at some time~$t$ can be represented by the use of a $Q$--symbol
representation.  However, the multidimensional domain of the
$Q$--symbol makes the visualization of this function challenging.  We
have suggested that the method of plotting slices of the $Q$--symbol is
desirable in this respect.  An example was provided where the dynamics
could be restricted to SU(2)$_{23}$, which makes the plots quite clear
and readily interpreted: a superposition of two SU(2) coherent states
is evident in these plots.

This approach may be extended to SU(n) coherent states by generalizing
the Hamiltonian~(\ref{H:nonlinear}) to $n$ interacting modes. However,
the SU(3) dynamics is the most relevant case to current physical
realizations.  Although nonlinear optics and Bose--Einstein
condensation have been specifically mentioned, any three--boson system
with the given nonlinear evolution~(\ref{H:nonlinear}) could yield superpositions of SU(3)
coherent states, provided that the initial state is itself an SU(3)
coherent state.

\acknowledgments{KN would like to thank G. J. Milburn for useful
discussions.  KN acknowledges the financial support of the Australian
International Education Foundation (AIEF).  This work has been
supported by an Australian Research Council Large Grant and by a
Macquarie University Research Grant.}

\end{document}